\documentclass[conference]{IEEEtran}
\IEEEoverridecommandlockouts
\usepackage{cite}
\usepackage{amsmath,amssymb,amsfonts}
\usepackage{algorithm}
\usepackage{algorithmic}
\usepackage{graphicx}
\usepackage{textcomp}
\usepackage{xcolor}
\usepackage{soul}
\usepackage{booktabs, threeparttable}
\usepackage{pifont}
\def\BibTeX{{\rm B\kern-.05em{\sc i\kern-.025em b}\kern-.08em
    T\kern-.1667em\lower.7ex\hbox{E}\kern-.125emX}}
\begin{document}

\title{ASMCap: An Approximate String Matching Accelerator for Genome Sequence Analysis Based on Capacitive Content Addressable Memory\vspace{-0.6em}}

\author{\IEEEauthorblockN{Hongtao Zhong$^1$, Zhonghao Chen$^1$, Wenqin Huangfu$^2$, Chen Wang$^1$, Yixin Xu$^3$, Tianyi Wang$^4$, Yao Yu$^4$,\\ Yongpan Liu$^1$, Vijaykrishnan Narayanan$^3$, Huazhong Yang$^1$, Xueqing Li$^{1\dagger}$}
$^1$BNRist/ICFC, Electronic Engineering Department, Tsinghua University, Beijing, China; $^2$Meta Platforms, Inc.;\\
$^3$Pennsylvania State University; $^4$Daimler Greater China Ltd.; $^{\dagger}$Email: xueqingli@tsinghua.edu.cn\vspace{-1.2em}
}
\maketitle
\begin{abstract}
Genome sequence analysis is a powerful tool in medical and scientific research. Considering the inevitable sequencing errors and genetic variations, approximate string matching (ASM) has been adopted in practice for genome sequencing. However, with exponentially increasing bio-data, ASM hardware acceleration is facing severe challenges in improving the throughput and energy efficiency with the accuracy constraint. 

This paper presents ASMCap, an ASM acceleration approach for genome sequence analysis with hardware-algorithm co-optimization. At the circuit level, ASMCap adopts charge-domain computing based on the capacitive multi-level content addressable memories (ML-CAMs), and outperforms the state-of-the-art ML-CAM-based ASM accelerators EDAM with higher accuracy and energy efficiency. ASMCap also has misjudgment correction capability with two proposed hardware-friendly strategies, namely the \textit{Hamming-Distance Aid Correction} (HDAC) for the substitution-dominant edits and the \textit{Threshold-Aware Sequence Rotation} (TASR) for the consecutive indels. Evaluation results show that ASMCap can achieve an average of 1.2x (from 74.7$\%$ to 87.6$\%$) and up to 1.8x (from 46.3$\%$ to 81.2$\%$) higher $F_1$ score (the key metric of accuracy), 1.4x speedup, and 10.8x energy efficiency improvement compared with EDAM. Compared with the other ASM accelerators, including ResMA based on the \textit{comparison matrix}, and SaVI based on the \textit{seeding} strategy, ASMCap achieves an average improvement of 174x and 61x speedup, and 8.7e3x and 943x higher energy efficiency, respectively.
\end{abstract}
\vspace{-0.3cm}
\section{Introduction}
\vspace{-0.2cm}
Recently, genome sequencing has received a lot of attention and triggered new innovations in several applications including precise medical care~\cite{guttmacher2010personalized}, virus surveillance~\cite{lu2020genomic}, evolutionary theory~\cite{prado2013great}, etc. With recent advances in sequencing technologies like Next-Generation Sequencing (NGS)~\cite{shendure2008next} or Third-Generation Sequencing (TGS)~\cite{schadt2010window}, massive amounts of genomics data can be generated at low cost. Such exponentially increasing bio-data scale much faster than the computing capability~\cite{canzar2015short}, which brings severe challenges to the genome sequence analysis.

In genome sequence analysis, small random fragments of the original DNA sequence extracted by machines, called \textit{reads}, are analyzed by a computational process called \textit{read mapping}. Specifically, each read is aligned to one or more possible locations in the reference sequence. Then, at these locations, matches and differences, i.e., \textit{distance}, are determined between the read and the reference sequence segments~\cite{alkan2009personalized}. Considering the inevitable errors caused by sequencing, genetic mutations and variations, Approximate String Matching (ASM) is necessary in the read mapping.
\begin{figure}[t]
  \centering
  \setlength{\abovecaptionskip}{-0.1cm}
  \includegraphics[width=8.2cm, page=1]{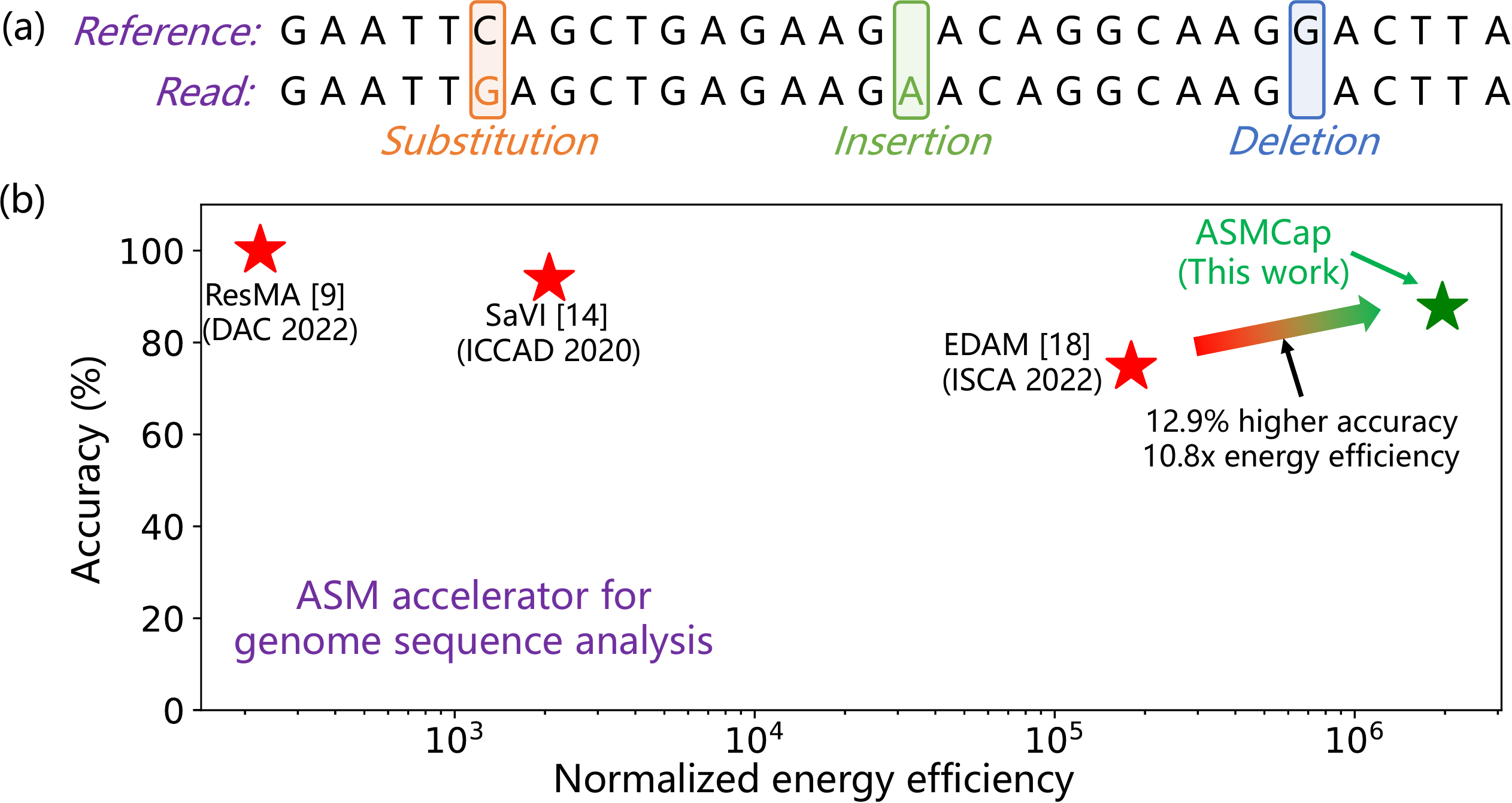}
  \caption{ASM acceleration for genome sequence analysis: (a) Three types of errors: substitution, insertion, and deletion; (b) Higher energy efficiency is desired for ASM acceleration.}
  \vspace{-0.6cm}
  \label{errors}
\end{figure}

In ASM, as shown in Fig.~\ref{errors}(a), in addition to substitutions, there are two basic errors: insertions and deletions (collectively called \textit{indels}). These three types of errors are typically referred to as \textit{edits}. In the von Neumann architecture, finding the \textit{Edit Distance} (ED), involves many complex operations and produces huge intermediate data with the growing sequence length, which further increases the data movement between the on-chip processors and the off-chip memory or storage~\cite{wei2017string}. It is a severe challenge towards efficient genome sequence analysis. 

ED can be calculated exactly using \textit{Comparison Matrix} (CM), but this iterative process is very costly even with the filtering and PIM techniques in ResMA~\cite{li2022resma}. Therefore, the \textit{seeding} strategy is frequently utilized in genomics~\cite{altschul1990basic, liu2016fast}, and several corresponding PIM-based accelerators have been developed on different platforms~\cite{huangfu2018radar, huangfu2020nest, laguna2020seed}. These two strategies find the most possible matched location by exactly matching all \textit{k}-length read fragments, (called \textit{k}-mers), with the reference and then \textit{extending} or \textit{voting} the matched \textit{k}-mers. Therefore, significant performance improvement can be achieved over the CM method, but they still suffer from limited throughput due to the computationally expensive \textit{k}-mer counting and the \textit{extending} or \textit{voting} process~\cite{laguna2020seed}.

Recently, EDAM, an ASM accelerator based on the Multi-Level Content Addressable Memories (ML-CAMs)~\cite{ni2019ferroelectric, garzon2022hamming, ma2022capcam}, was reported in~\cite{hanhan2022edam} with orders of magnitudes higher energy efficiency, as shown in Fig.~\ref{errors}(b). Unlike the conventional CAMs or ML-CAMs, for a certain base, EDAM matches not only the co-located base but also its left and right neighbors. Therefore, it can tolerate \textit{intra-mer} edits, making it possible to directly match the reads with the reference without fragmentation. Combining the \textit{in-situ} parallel match of CAMs, EDAM achieves much higher performance. However, the current-mode computing and sensing in EDAM are not variation-resistant and scalable, which limits the read length and degrades energy efficiency. Furthermore, the matching method in EDAM may induce misjudgments under certain conditions, which affects the accuracy and limits its application.

From the analysis above, as shown in Fig.~\ref{errors}(b), there is a trade-off between accuracy and energy efficiency in existing ASM accelerators, and it is very challenging to find a good balance. In this paper, we propose ASMCap, a novel ASM architecture based on the capacitive ML-CAMs~\cite{ma2022capcam}. ASMCap investigates the matching method proposed in EDAM and makes significant progress: (i) The charge-domain computation with no area cost is proposed to provide linear and stable voltage output scaled by the matching results, and achieve much higher reliability and read length upper bound with lower capacitor variations; (ii) ASMCap proposes two hardware-friendly heuristic misjudgment correction strategies, namely the \textit{Hamming-Distance Aid Correction} (HDAC) and the \textit{Threshold-Aware Sequence Rotation} (TASR), to improve the accuracy with negligible area overheads.

The contributions of this paper are as follows:
\begin{itemize}
    \item We propose ASMCap, a novel ASM architecture based on the capacitive ML-CAMs with inherent high reliability, throughput, and energy efficiency.
    \item We propose two hardware-friendly heuristic misjudgment correction strategies, i.e., HDAC and TASR, for higher accuracy with negligible overheads.
    \item We perform extensive experiments for ASMCap. Results show an average of 1.2x and up to 1.8x higher $F_1$ accuracy score than EDAM, and an average of 174x, 61x and 1.4x speedup and 8.7e3x and 943x and 10.8x higher energy efficiency than the state-of-the-art ASM accelerators ReSMA, SaVI and EDAM, respectively.
\end{itemize}

Next, Section II introduces the genome sequence analysis background. Section III presents ASMCap with further software optimizations in Section IV. Section V shows the experimental results and Section VI concludes this work.
\vspace{-0.4cm}
\section{Background}
\vspace{-0.2cm}
\subsection{Genome Sequence Analysis}
\vspace{-0.1cm}

Genome sequences consist of four types of bases: Adenine (A), Guanine (G), Cytosine (C), and Thymine (T). A-T and C-G are complementary base pairs (bps). To perform the genome sequence analysis, the positions of the reads need to be determined, and read mapping is to align each read sequence to one or more possible locations in the reference sequence. The \textit{read alignment} includes \textit{Exact Alignment}~\cite{li2009fast} and \textit{Approximate Alignment}~\cite{altschul1990basic,cali2020genasm}. Considering the inevitable errors, \textit{Approximate Alignment}, i.e., ASM, needs to be supported in practical genome sequencing~\cite{cali2020genasm}.
\begin{figure}[t]
  \centering
  \setlength{\abovecaptionskip}{-0.1cm}
  \includegraphics[width=7.6cm, page=2]{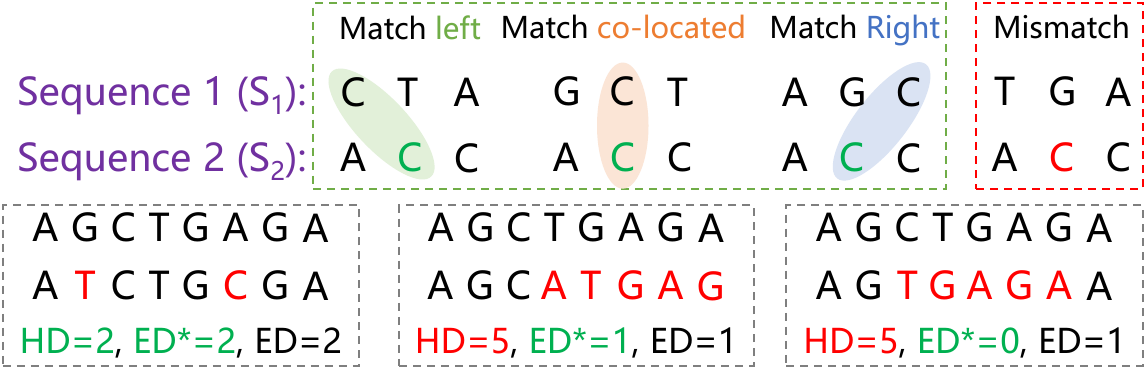}
  \caption{The adopted matching method in ASMCap~\cite{hanhan2022edam}.}
  \label{EDAM}
  \vspace{-0.4cm}
\end{figure}
\begin{figure}[t]
  \centering
  \setlength{\abovecaptionskip}{-0.1cm}
  \includegraphics[width=\linewidth, page=3]{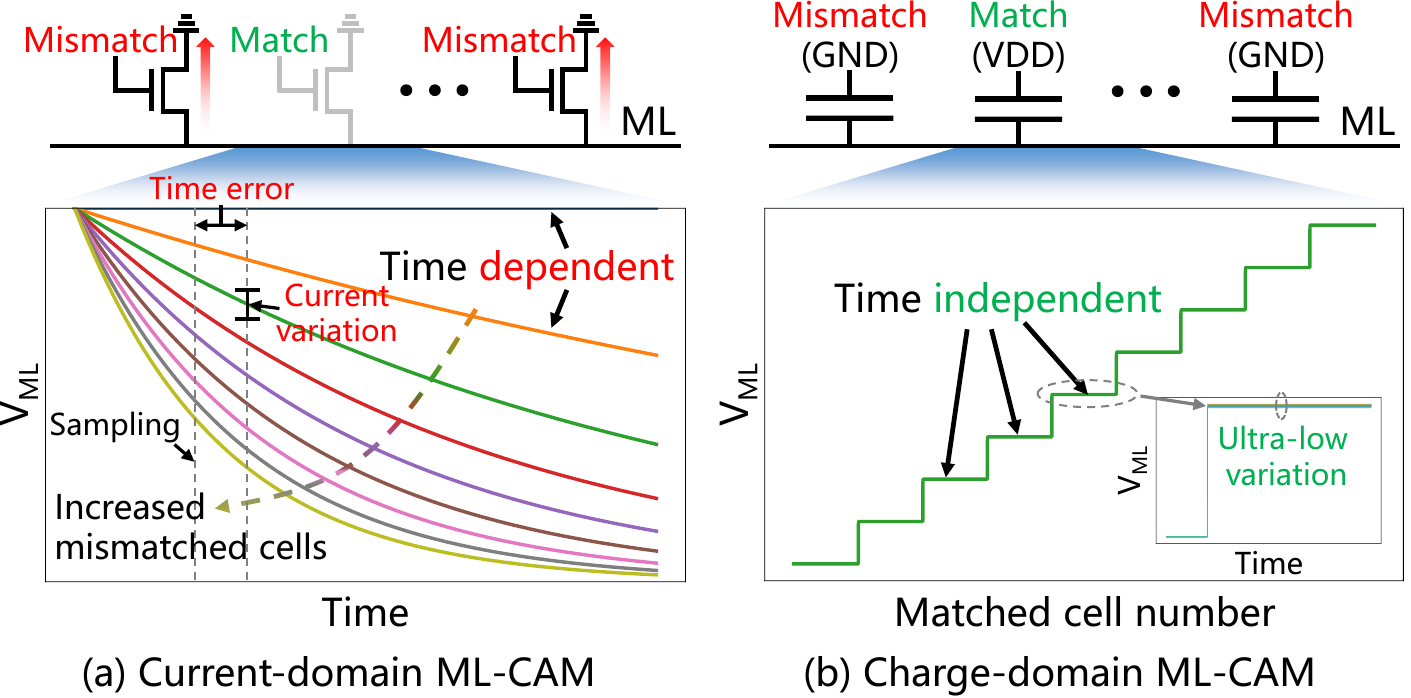}
  \caption{ML-CAMs in the (a) current-domain~\cite{ni2019ferroelectric, garzon2022hamming} and (b) charge-domain~\cite{ma2022capcam} for multi-level content matching.}
  \label{current-charge}
  \vspace{-0.6cm}
\end{figure}

There are two major error sources in genome sequences: sequencing errors and genetic variations. For example, NGS~\cite{shendure2008next} can produce \textit{short reads} ($\sim$50-500 bps) with \textit{low errors} ($\sim$0.1\%-1\%)~\cite{goodwin2016coming}, while TGS~\cite{schadt2010window} can produce \textit{long reads} (thousands to millions of bps) with \textit{high errors} ($\sim$10\%-15\%)~\cite{cali2020genasm}. The error rate of genetic variations is relatively low ($\sim$0.1\% for human~\cite{canzar2015short}).
\vspace{-0.2cm}
\subsection{Approximate String Matching and Related Works~\label{ASM background}}\vspace{-0.1cm}
ED between two sequences is the minimum number of edits required to transform one sequence to the other one. Given a read sequence $R$=$[r_1r_2...r_m]$, a reference sequence $Q$=$[q_1q_2...q_n]$,  \textit{m}=$|R|$, \textit{n}=$|Q|$, \textit{m}$\leq$\textit{n} and a ED threshold \textit{T}, the ASM goal is to find all possible subsequences $S$ in $Q$ and their positions such that ED($S$, $R$)$\leq$\textit{T}. 

ED can be calculated exactly using the \textit{Comparison Matrix} $M[i,j]$, but the time complexity of this ED computation is $\mathcal{O}(N^2)$, which is very costly~\cite{li2022resma}. ReSMA~\cite{li2022resma}, a Resistive Random Access Memory (RRAM) based PIM accelerator, utilizes the RRAM crossbars to exploit anti-diagonal parallelism in $M[i,j]$ and the RRAM CAMs for filtering, which gains great performance improvement. However, the iterative process during CM computation incurs massive intermediate data and updates the crossbars frequently, which heavily degrades the performance and energy efficiency.
\begin{figure*}[t]
  \centering
  \setlength{\abovecaptionskip}{-0.2cm}
  \includegraphics[width=\linewidth, page=4]{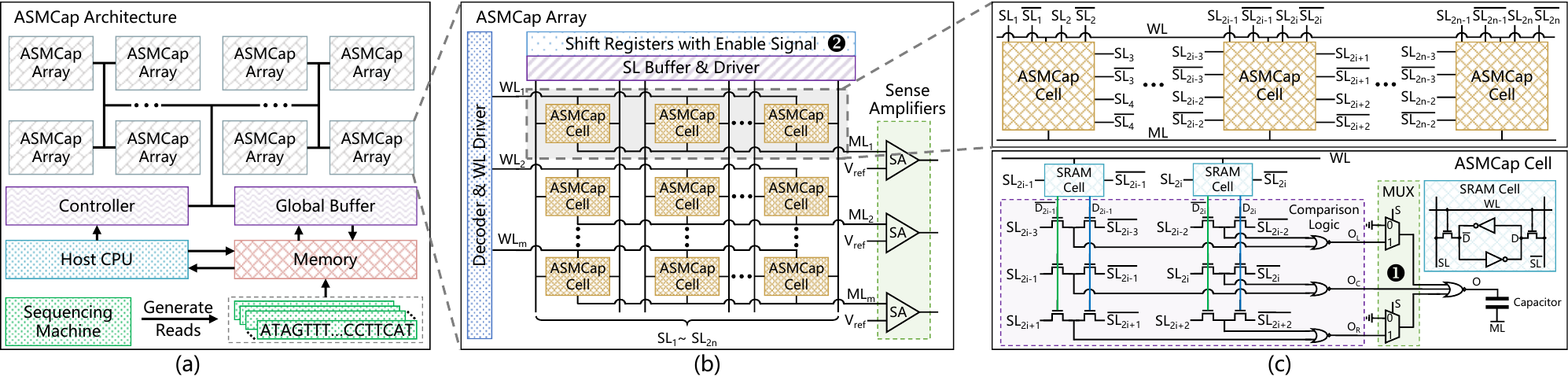}
  \caption{The proposed ASMCap architecture. (a) Top architecture; (b) ASM array in the ASMCap units; (c) Detailed circuit design in an ASMCap cell.}
  \vspace{-0.6cm}
  \label{arch}
\end{figure*}

A more common method in genomics is the \textit{seeding} strategy, including the \textit{seed-and-extend} strategy~\cite{altschul1990basic} and the \textit{seed-and-vote} strategy~\cite{liu2016fast}. The \textit{seeding} strategy splits the read into many \textit{k}-mers, and then finds all locations of the reference for each \textit{k}-mer by \textit{exact matching}, and finally, extends or votes the matched \textit{k}-mers to find the best possible match between the read and the reference. The \textit{voting} process is faster than the \textit{extending} process but suffers from the accuracy loss~\cite{liu2016fast}. 

There have been several existing accelerators for the \textit{seeding} strategy based on 3D RRAM, DIMM, TCAMs, etc. ~\cite{huangfu2018radar, huangfu2020nest, laguna2020seed}. These two strategies can both be much faster than the CM computing, but still suffer from limited throughput due to the memory-hungry and time-consuming \textit{k-mer} counting~\cite{huangfu2020nest}. In addition, both \textit{extending} and \textit{voting} processes involve complex computation, which further degrades the performance. 

The recent ASM accelerator EDAM~\cite{hanhan2022edam} is based on ML-CAMs~\cite{ma2022capcam, ni2019ferroelectric, garzon2022hamming}. As illustrated in Fig.~\ref{EDAM}, EDAM performs the search operation for a certain base with not only the co-located base but also the left and right neighbors of the co-located base. If there is at least one match, the matching result of this base is \textit{`match'}. Otherwise, the result is \textit{`mismatch'}. The distance estimated by EDAM is defined as ED* in this paper. From the examples in Fig.~\ref{EDAM}, it is seen that ED* can be more close to ED compared with \textit{Hamming Distance} (HD) when insertions or deletions occur. In other words, EDAM can tolerate \textit{intra-mer} edits, support much larger $k$, and even match the read directly without fragmentation. In this way, the \textit{k}-mer counting and the \textit{extending} or \textit{voting} process can be reduced and even eliminated. With the \textit{in-situ} parallel match of CAMs, EDAM can achieve ultra-high throughput.
\vspace{-0.2cm}
\subsection{Current Domain and Charge Domain ML-CAMs}
\vspace{-0.1cm}
ML-CAM is one type of CAM that can not only perform \textit{exact matching or not}, but also detect the \textit{degree of match}. EDAM uses the current-domain ML-CAM~\cite{ni2019ferroelectric}, as shown in Fig.~\ref{current-charge}(a). The matching result of each base controls a transistor: a \textit{`match'} turns off the transistor, and a \textit{`mismatch'} turns on the transistor. The matchline (\textit{ML}) is pre-charged, and the decreasing slope of the \textit{ML} voltage $V_{ML}$ scales with the number of mismatched cells. However, this dynamic timing-dependent sensing method is inherently vulnerable to device and timing-control variations and power-consuming, which results in poor scalability and limits the maximum read length that EDAM can support. Besides, the pre-charge operation in EDAM also consumes significant power.

To address the inherent challenge of the current-domain ML-CAM, charge-domain ML-CAM, i.e., the capacitive ML-CAM, was proposed~\cite{ma2022capcam} for one-shot learning. As shown in Fig.~\ref{current-charge}(b), the matching result of each base outputs a certain voltage (\textit{VDD} for \textit{`match'} and \textit{GND} for \textit{`mismatch'}) to the bottom plate of a capacitor, and the top plates of all capacitors are connected to \textit{ML}. It shows that $V_{ML}$ scales linearly with the degree of match, and is stable and timing-independent. Besides, capacitor variations are much less than the current variations, which enhances the accuracy and scalability.
\vspace{-0.2cm}
\section{Proposed ASMCap Architecture~\label{architecture}}
This section presents the proposed ASMCap, including the top architecture and detailed building blocks. Operation mechanisms and performance analysis are also included.
\vspace{-0.1cm}
\subsection{Top Architecture}
\vspace{-0.1cm}
Fig.~\ref{arch}(a) shows the top architecture of ASMCap. The reads generated by the sequencing machine are firstly stored in the memory and the global buffer can fetch the entire reads or \textit{k}-mers for the subsequent match according to the read length. The controller receives instructions from the host CPU and controls the process. The reads or \textit{k}-mers from the global buffer are sent to the ASMCap arrays through the H-tree.
\vspace{-0.1cm}
\subsection{ASMCap Array}
\vspace{-0.1cm}
As shown in Fig.~\ref{arch}(b), each ASMCap array includes \textit{M}$\times$\textit{N} ASMCap cells, the decoder, the wordline (\textit{WL}) driver, the searchline (\textit{SL}) buffer and driver, the sense amplifiers (SAs) on the matchline (\textit{ML}), and the shift registers with an enable signal. Each row stores a reference segment of the same length as the input reads or \textit{k}-mers. The decoder and the \textit{WL} driver obtain the address and select rows to execute write and search operations. \textit{SL} buffer and driver transform the input data to the corresponding voltage and drive the \textit{SLs} and $\overline{SL}$\textit{s}.

The SAs compare $V_{ML}$ with the reference voltage ($V_{ref}$). Different from EDAM~\cite{hanhan2022edam}, the sample and hold circuit and the timing are not required in ASMCap, thanks to the stable output voltage on the \textit{ML}, which reduces the search latency. the SAs are required to output  `1' (\textit{`match'}) when $V_{ML}\leq V_{ref}$ corresponding to ED*$\leq$\textit{T}. 

The shift registers with enable signal can rotate the input reads or \textit{k}-mers left or right base-by-base, and are utilized in the proposed TASR strategy in Section~\ref{TASR_method}.
\vspace{-0.1cm}
\subsection{ASMCap Cell}
\vspace{-0.1cm}
Fig.~\ref{arch}(c) shows the circuit design of the $i^{th}$ ASMCap cell in a row of the ASMCap array. The $i^{th}$ base of the reference segment is stored in the two 6T SRAM cells. The comparison logic compares the stored base with the co-located base and its left and right neighbors in the read, and produces the partial matching results, i.e., $O_C$, $O_L$, and $O_R$, respectively. If match occurs, the corresponding partial matching result is `1'. Then, two multiplexers (MUXs) control the matching mode. If the select signal of the MUXs (\textit{S}) is `1', the output matching result (\textit{O}) is $\overline{O_C+O_L+O_R}$. Otherwise, $O=\overline{O_C}$. In this way, the ASMCap array can perform the search operation using both ED* and HD, which is utilized in the proposed HDAC strategy in Section~\ref{HDAC_method}. Note that the select signal \textit{S} is shared by all MUXs in the ASMCap array, and \textit{MLs} are not required to be pre-charged during the search operation as EDAM~\cite{hanhan2022edam}, which also reduces the search latency.

For the matching mode using ED*, if at least one partial matching result is `1', \textit{O} is `0' and the $i^{th}$ cell is a matched cell. Otherwise, \textit{O} is `1', and the $i^{th}$ cell is a mismatched cell. Therefore, $V_{ML}=\frac{n_{mis}}{N}VDD$, where $n_{mis}$ is the total number of mismatched cells equivalent to ED*. Therefore, ED*$\leq$\textit{T} corresponds to $V_{ML}\leq V_{ref}$ when $V_{ref}$ is set to $\frac{T}{N}VDD$. 

Furthermore, if all capacitors in the ASMCap array follow the independent and identically distributed normal distribution $\mathcal{N}(\mu_C,\,\sigma^2_C)$, as~\cite{ma2022capcam}, the energy consumption per search operation ($E_S$) and the $V_{ML}$ variance can be estimated as:
\begin{equation}
    \vspace{-0.1cm}
    E_S\approx\frac{Mn_{mis}(N-n_{mis})}{N}\mu_CVDD^2
\end{equation}
\begin{equation}
    \label{variation}
    Var(V_{ML})\approx\frac{n_{mis}(N-n_{mis})}{N^3}(\frac{\sigma_C}{\mu_C})^2VDD^2
    \vspace{-0.1cm}
\end{equation}
It is seen that $E_S$ and $Var(V_{ML})$ are small when $n_{mis}$ is close to 0 or \textit{N}. Considering the genome sequence characteristics, $n_{mis}$ is close to \textit{N} for most rows, making ASMCap superior to EDAM with much lower power consumption and variations.
\vspace{-0cm}
\section{More Software Optimizations}
Although EDAM can tolerate edits and is efficient in estimating ED when the indels occur, misjudgments limit the accuracy of EDAM. Because of the huge design space, it is challenging to correct those misjudgments with low hardware overhead. This section focuses on two common kinds of misjudgments and describes two proposed corresponding hardware-friendly heuristic strategies, i.e., HDAC and TASR, for misjudgment correction with the overhead analysis.
\vspace{-0.1cm}
\subsection{Hamming-Distance Aid Correction~\label{HDAC_method}}
\begin{figure}[t]
  \centering
  \setlength{\abovecaptionskip}{-0.1cm}
  \includegraphics[width=8cm, page=5]{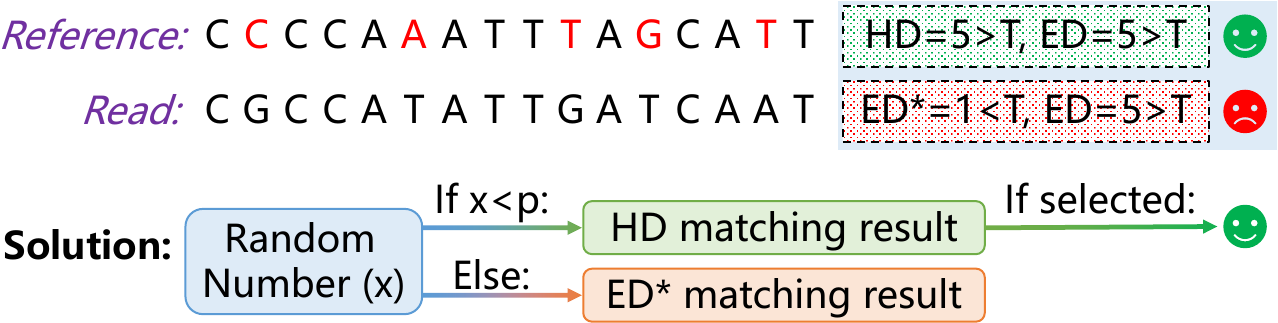}
  \caption{The proposed HDAC strategy for substitution-dominant edits (5 substitutions, no indels, and $T=4$ in this example).}
  \label{HDAC}
  \vspace{-0.2cm}
\end{figure}

\textbf{Observation:} As illustrated in Fig.~\ref{HDAC}, a common kind of misjudgment happens when there are many substitutions but few or even no indels. If performing the search operation using HD, most edits can be discovered. However, EDAM performs the search operation for a certain base with the left and right neighbors, many edits are likely to be hidden. It leads to the misjudgment, i.e., False Positive (FP), when the threshold \textit{T} is between ED* and ED (The ED* matching result is \textit{`match'} but the result should be \textit{`mismatch'} due to ED$>$\textit{T}).

\textbf{Solution:} We propose the HDAC strategy to address this issue. As shown in Fig.~\ref{HDAC}, we perform the search operation using both HD and ED*, and then select the HD matching result with a \textit{p} possibility. The pseudo codes of the HDAC are listed in Algorithm~\ref{alg:HDAC}. The main challenge is how to design the function $f()$ in \textbf{\textit{line 1}} such that \textit{p} is large enough when edits are mainly from substitutions, and decreases rapidly when more indels occur. In this paper, the function is designed as $\frac{e_s}{e_s+e_{id}}e^{-(\alpha e_{id}+\beta T)}$, where $e_s$ is the substitution error rate, $e_{id}$ is the indel error rate ($e_{id}=e_i+e_d$, where $e_i$ is the insertion error rate, and $e_d$ is the deletion error rate), and $\alpha$ and $\beta$ are constants. The $\frac{e_s}{e_s+e_{id}}$ part is to enlarge \textit{p} with larger proportion of substitutions in edits. The $e^{-\alpha e_{id}}$ part is to exponentially alleviate the effect of more indels due to larger $e_{id}$. The $e^{-\beta T}$ part is to exponentially alleviate the effect of a larger threshold $T$ because many matching results of indel-induced large HD are \textit{`mismatch'}, but more of them with small ED should be \textit{`match'} with larger \textit{T}, which leads to another misjudgment, i.e., False Negative (FN). Besides, $\frac{e_s}{e_s+e_{id}}e^{-(\alpha e_{id}+\beta T)}$ can be pre-processed off-line to reduce the computation overhead. It is also necessary to point out that this function is only an example. More accurate functions can be designed analytically or obtained by other powerful tools such as neural networks.
\newlength{\textfloatsepsave}
\setlength{\textfloatsepsave}{\textfloatsep} 
\setlength{\textfloatsep}{0pt} 
\begin{algorithm}[t]
    \caption{Hamming-Distance Aid Correction}
    \label{alg:HDAC}
    \renewcommand{\algorithmicrequire}{\textbf{Input:}}
    \renewcommand{\algorithmicensure}{\textbf{Output:}}
    \begin{algorithmic}[1]
        \REQUIRE $O_{HD}$: The HD matching result; $O_{ED}*$: The ED* matching result; $e_s$: The substitution error rate; $e_{id}$: The indel error rate; $T$: The threshold %%input
        \ENSURE $O$: The final matching result   %%output
        \STATE Let $p \leftarrow f(e_s,e_{id},T)$; \textcolor{blue}{\textit{// Can be pre-processed off-line}}
        \IF{$O_{HD} \neq O_{ED}*$}
            \STATE Generate a random number $\mathbb{X} \sim U(0,1)$;
            \IF{$\mathbb{X} < p$}
                \STATE Let $O \leftarrow O_{HD}$;
            \ELSE
                \STATE Let $O \leftarrow O_{ED}*$;
            \ENDIF
        \ENDIF
    \end{algorithmic}
\end{algorithm}

\textbf{Overhead Analysis:} HDAC involves two additional MUXs per ASMCap cell in~\ding{182} of Fig.~\ref{arch}. The MUXs are achieved by two NMOS transistors controlled by \textit{S} and $\Bar{S}$, respectively. Estimated from the layout, two MUXs induce about 0.1$\%$ area overhead through layout optimization. Besides, the HDAC strategy also induces one more cycle to perform the search operation using HD, we can disable the HDAC strategy when \textit{p} is less than a certain threshold (e.g., 1$\%$).
\vspace{-0.1cm}
\subsection{Threshold-Aware Sequence Rotation~\label{TASR_method}}
\begin{figure}[t]
  \centering
  \setlength{\abovecaptionskip}{0cm}
  \includegraphics[width=7.4cm, page=6]{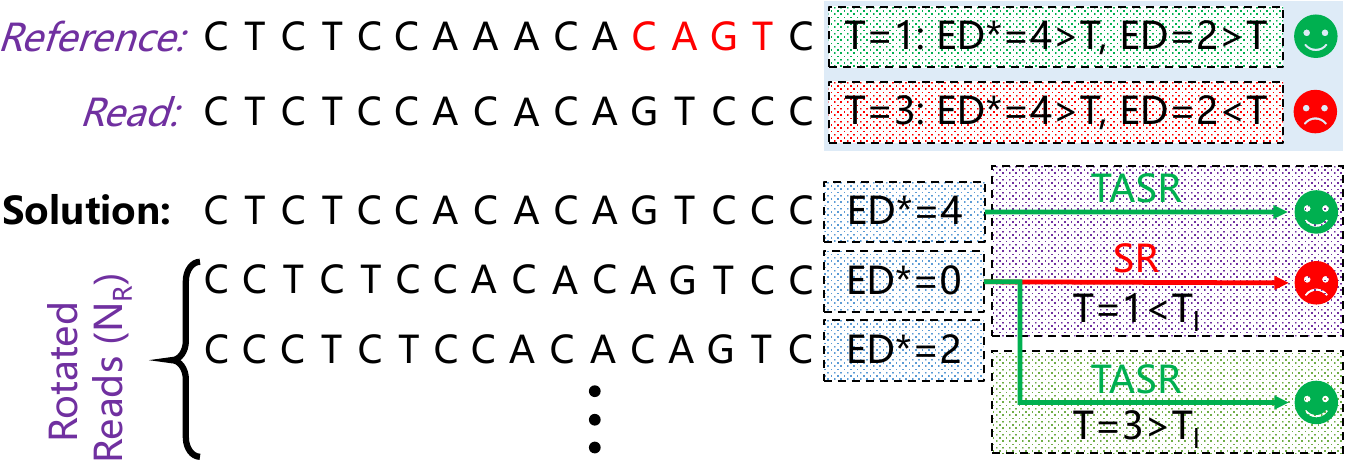}
  \caption{The proposed TASR strategy for consecutive insertions or deletions (delete consecutive `AA' in the read and $T_l=2$ in this example).}
  \label{TASR}
  \vspace{-0.2cm}
\end{figure}

\textbf{Observation:} As illustrated in Fig.~\ref{TASR}, another common kind of misjudgment happens when several consecutive insertions or deletions occur, which causes ED* to be much larger than ED. If \textit{T} is between ED and ED*, the FN happens.

\textbf{Solution:} An effective way is Sequence Rotation (SR)~\cite{hanhan2022edam}. SR rotates the read base-by-base $N_R$ times (can be the left or right rotation), and the reference segment is compared with the original read and $N_R$ rotated reads. If there is at least one \textit{`match'}, the final matching result is \textit{`match'}. However, as illustrated in Fig.~\ref{TASR}, the ED*s between the reference segment and some rotated reads may be smaller than ED, so the FP occurs if \textit{T} is between ED and these ED*s, especially when \textit{T} is relatively small. In this paper, we propose an improved strategy, i.e., TASR, and the pseudo codes are listed in Algorithm~\ref{alg:TASR}. We set a lower bound of \textit{T}, called $T_l$, such that the rotations are triggered only if $T\geq T_l$. In this way, the FP is avoided when \textit{T} is relatively small. $T_l$ also involves huge design space exploration, and here we design $T_l$ as $\lceil\frac{\gamma}{e_{id}}m\rceil$, where $\gamma$ is a constant and \textit{m} is the read length, considering $T_l$ should be small when $e_{id}$ is high to improve accuracy and be large when $e_{id}$ is low to save time and power consumption.

\textbf{Overhead:} TASR requires additional shift registers with enable signals in~\ding{183} of Fig.~\ref{arch}, but the average area overhead per cell is only $0.2\%$. Besides, the \textit{rotation-and-comparison} process also induces $N_R$ more cycles. With the $T_l$ limitation, the average latency overhead can be significantly reduced.

\vspace{-0.2cm}
\section{Experiments and Discussion}
\vspace{-0.3cm}

\begin{algorithm}[t]
    \caption{Threshold-Aware Sequence Rotation}
    \setlength{\abovecaptionskip}{0cm}
    \label{alg:TASR}
    \renewcommand{\algorithmicrequire}{\textbf{Input:}}
    \renewcommand{\algorithmicensure}{\textbf{Output:}}
    \begin{algorithmic}[2]
    \REQUIRE $R$: The read; $S$: The reference segment; $N_R$: The total rotation number; $T$: The threshold; $T_{l}$: The lower bound of $T$ to trigger the sequence rotation%%input
    \ENSURE $O$: The final matching result  %%output
    \IF{$T < T_{l}$}
        \STATE Let $O \leftarrow$ ED*($S$, $R$) $\leq T$;
    \ELSE
        \FOR{$i$ from 0 to $N_R$}
            \STATE Let $R_i \leftarrow Rotate(R, i)$; \textcolor{blue}{\textit{// Rotate left (right) $i$ bases}}
            \STATE Let $O \leftarrow$ $O$ \textbf{or} ED*($S$, $R_i$) $\leq T$;
        \ENDFOR
    \ENDIF
    \end{algorithmic}
\end{algorithm}
\vspace{-0.1cm}
\subsection{Experimental Setup}
\vspace{-0.1cm}
\textbf{Genome Sequence Analysis:} The datasets used in the experiments are the human genome from NCBI~\cite{NCBI}. In this paper, we mainly focus on \textit{short reads} considering the capacity and the sensing ability of the ML-CAMs. The reads are set to 256-base length within the typical range of the read length ($\sim$50-500 bps~\cite{goodwin2016coming}), and extracted from random positions in human DNA sequences. Then, edits are randomly injected, which creates metagenomic datasets with the following two kinds of mixed error rates for edits considering the typical error range of the \textit{short reads} ($\sim$0.1\%-1\%)~\cite{goodwin2016coming}:
\begin{itemize}
    \item \textbf{Condition A:} $e_s$ = $1\%$ and $e_i$ = $e_d$ = $0.05\%$;
    \item \textbf{Condition B:} $e_s$ = $0.1\%$ $e_i$ = $e_d$ = $0.5\%$;
\end{itemize}

Each 256-base-long read in the metagenomic datasets is directly sent to the EDAM and ASMCap arrays without fragmentation. The references extracted from the human genome are segmented and stored in the EDAM and ASMCap arrays. 

To compare the accuracy, we evaluate the $F_1$ score as EDAM, and the $F_1$ score is calculated by:
\vspace{-0.2cm}
\begin{equation}
    Sensitivity=\frac{TP}{TP+FN}, Precision=\frac{TP}{TP+FP}
\end{equation}
\begin{equation}
    F_1=\frac{2 \times Sensitivity \times Precision}{Sensitivity+Precision}
    \vspace{-0.2cm}
\end{equation}
where the TP is True Positive that the matching result is \textit{`match'} and it indeed should be \textit{`match'}. Besides, as EDAM, we use the popular tool Kraken2~\cite{wood2019improved} as a baseline and the normalized $F_1$ score is normalized by $F_1(Kraken2)$.

\textbf{EDAM and ASMCap Configurations:} ASMCap is designed in a commercial 65nm CMOS process, and 2fF MIM capacitors are used in ASMCap cells. The search voltage, the array size, and the array number of EDAM and ASMCap are both set to 1.2V, 256$\times$256, and 512, respectively. For the proposed two strategies of ASMCap, $\alpha$ and $\beta$ are set to 200 and 0.5 in HDAC, respectively, and $N_R$ and $\gamma$ are set to 2 and $2\times10^{-4}$ in TASR, respectively.
\begin{table}[t]
    \centering
    \setlength{\abovecaptionskip}{0cm}
    \caption{Circuit-level Comparison between ASMCap and EDAM}
    \label{comparison}
    \begin{threeparttable}
    \begin{tabular}{ccc}
    \toprule
         &  EDAM~\cite{hanhan2022edam} & ASMCap\\
    \midrule
       ML-CAM Mode  & Current domain & Charge domain\\
       Technology & 65nm & 65nm\\
       Cell Area & 33.4$\mu m^2$ (1.4x) & 24.0$\mu m^2$ (1x)\\
       Supply voltage & 1.2V & 1.2V \\
       Search time & 2.4ns (2.6x) & 0.9ns (1x) \\
       Average Power per cell\tnote{1} & 1.0$\mu$W (8.5x) & 0.12$\mu$W (1x) \\
    \bottomrule
    \end{tabular}
    \begin{tablenotes}
        \item[1] The average power under two conditions.
    \end{tablenotes}
    \end{threeparttable}
    \vspace{-0.6cm}
\end{table}

\textbf{Other ASM Solutions:} We compare ASMCap with other state-of-the-art ASM solutions mentioned in Section~\ref{ASM background}, including the CM computation using i9-10980XE CPU (CM-CPU) as the baseline and the CM computation using RRAM-based PIM accelerators (ReSMA~\cite{li2022resma}) and the \textit{seed-and-vote} strategy using TCAM (SaVI~\cite{laguna2020seed}). All these ASM solutions also process 256-base-long reads for a fair comparison.
\vspace{-0.1cm}
\subsection{Area and Power Breakdown of ASMCap}
\vspace{-0.1cm}
For a 256$\times$256 ASMCap array, the area and power are 1.58$mm^2$, and 7.67mW, respectively. For area, more than 99$\%$ of the area is occupied by the ASMCap cells. For power, the ASMCap cells, the shift registers, and SAs occupy 75$\%$, 19$\%$, and 6$\%$ of power, respectively.
\setlength{\textfloatsep}{\textfloatsepsave}
\subsection{Circuit-Level Comparison with EDAM}
\vspace{-0.1cm}
We evaluate the circuit-level simulations using Cadence Virtuoso, and Table~\ref{comparison} lists the circuit-level comparison between ASMCap and EDAM. Results show that ASMCap achieves 1.4x cell area reduction, 2.6x less search time, and 8.5x power reduction compared with EDAM. 

Through layout estimation, the area reduction results from both the reduced transistors for the discharge process in EDAM and layout optimizations. Although the area of a 65nm 2fF MIM capacitor is about 1.4$\mu m^2$~\cite{ma2022capcam}, the capacitors can be placed on top of the cell, which induces no area penalty considering the large cell area.

The less search time is from the skipped pre-charge operation and the sampling operation in EDAM, as analyzed in Section~\ref{architecture}. Besides, the smaller parasitic effect due to the smaller area also helps ASMCap reduce the search time. The power reduction is mainly from the charge-domain computation mode, as analyzed in Section~\ref{architecture}.

\begin{figure}[t]
  \centering
  \setlength{\abovecaptionskip}{0cm}
  \includegraphics[width=8.0cm]{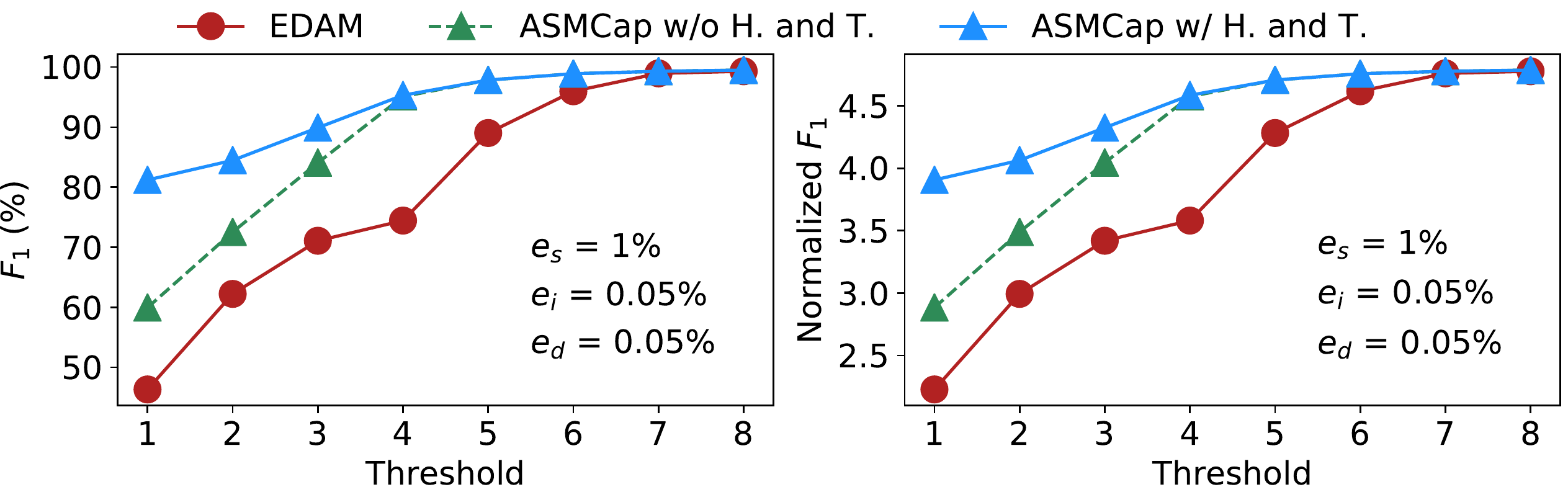}
  \includegraphics[width=8.0cm]{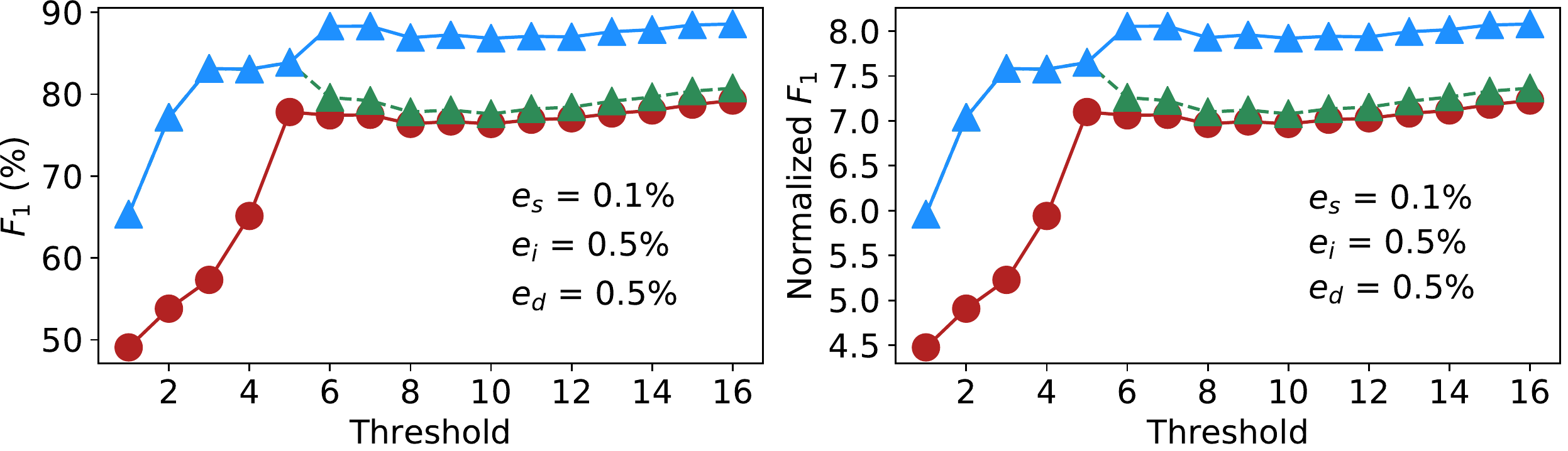}
  \caption{Accuracy ($F_1$) comparison between ASMCap with the proposed strategies and EDAM. H. and T. represent HDAC and TASR, respectively.}
  \label{hs}
  \vspace{-0.6cm}
\end{figure}

\vspace{-0.1cm}
\subsection{Accuracy Comparison with EDAM}
\vspace{-0.1cm}
To perform the accuracy comparison, Monte Carlo simulations are carried out for both ASMCap and EDAM. The estimated current variation in EDAM is 2.5$\%$, supporting at most 44 distinguishable states under the $3\sigma$ constraint for $V_{ML}$. Meanwhile, the capacitor variation in ASMCap is only 1.4$\%$. Combined with Equation~\ref{variation}, ASMCap can support 566 distinguishable states even with the worst case.

In this way, as shown in Fig.~\ref{hs}, ASMCap without the HDAC and TASR strategies achieves an average of 1.13x and 1.11x higher $F_1$ in Condition A and B, respectively (1.12x higher $F_1$ on average). Furthermore, the proposed HDAC strategy achieves an average of 1.07x further higher $F_1$ in Condition A, and the proposed TASR strategy achieves an average of 1.08x further higher $F_1$ in Condition B. Therefore, ASMCap with the HDAC and TASR strategies achieves an average of 1.2x higher $F_1$ (from 74.7$\%$ to 87.6$\%$), and up to 1.8x higher $F_1$ (from 46.3$\%$ to 81.2$\%$) when \textit{T}=1 in Condition A. Finally, compared with Kraken~\cite{wood2019improved} with \textit{exact matching}, ASMCap can achieve an average of 4.5x and 7.7x higher $F_1$ in Condition A and B, respectively (6.6x higher $F_1$ on average).
\vspace{-0.2cm}
\subsection{Comparison with Existing ASM Accelerators}
\begin{figure}[t]
  \centering
  \setlength{\abovecaptionskip}{-0.2cm}
  \includegraphics[width=\linewidth]{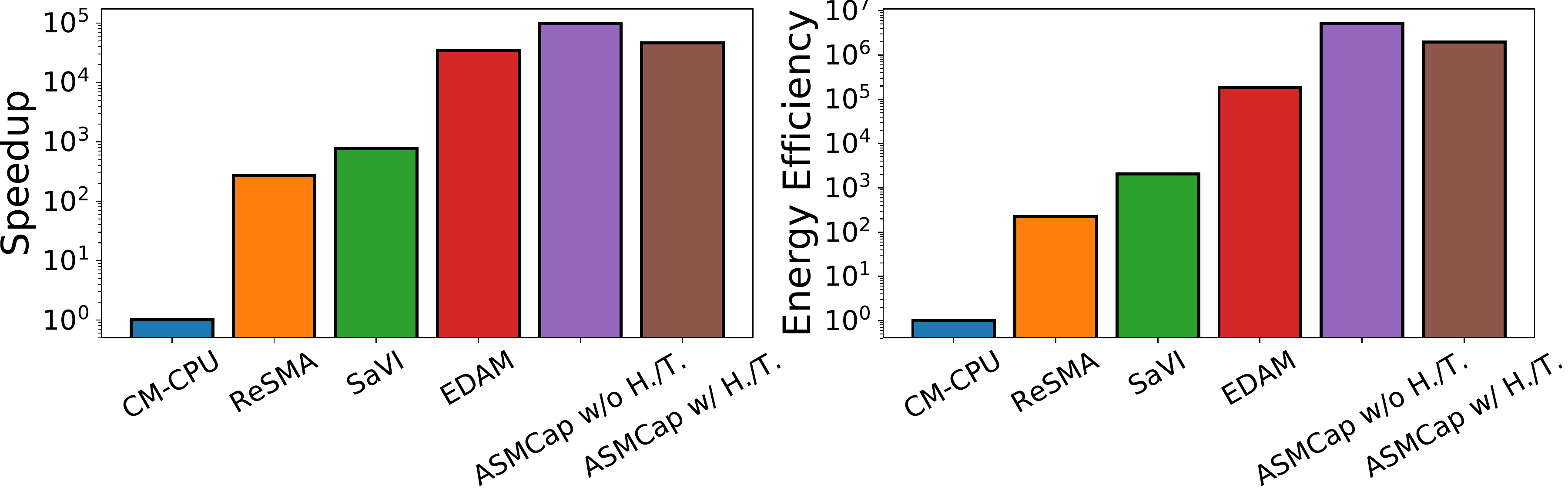}
  \caption{Speedup and energy efficiency comparison between ASMCap and existing ASM accelerators. H. and T. represent HDAC and TASR, respectively.}
  \label{speed}
  \vspace{-0.6cm}
\end{figure}

\vspace{-0.1cm}
To evaluate the system-level performance and energy efficiency, 512 ASMCap arrays are utilized with 64Mb memory capacity, which can entirely store some small virus sequences (e.g., SARS-CoV-2 in coronavirus pandemic~\cite{hanhan2022edam}). As shown in Fig.~\ref{speed}, without the proposed HDAC and TASR strategies, ASMCap achieves an average of 9.7e4x, 362x, 126x and 2.8x speedup and an average of 5.1e6x, 2.3e4x, 2.4e3x and 28x energy efficiency compared with CM-CPU, ReSMA, SaVI, and EDAM, respectively. Considering the average effect of the proposed HDAC and TASR strategies, ASMCap achieves an average of 4.7e4x, 174x, 61x and 1.4x speedup and an average of 2.0e6x, 8.7e3x, 943x and 10.8x energy efficiency compared with CM-CPU, ReSMA, SaVI, and EDAM, respectively.

For accuracy, the CM computation can reach 100$\%$ accuracy, while the \textit{seed-and-vote} strategy achieves about 93.8$\%$ accuracy on average~\cite{liu2016fast}. Therefore, with ultra-high throughput and energy efficiency with comparable accuracy, ASMCap is more suitable for the task-intensive but accuracy-insensitive scenarios such as fast testing.
\vspace{-0.2cm}
\section{Conclusion}
\vspace{-0.1cm}
This paper proposes ASMCap, an ASM acceleration approach based on capacitive ML-CAMs that achieves the highest reported energy efficiency. Compared with prior ML-CAM based accelerators, the charge-domain computation of ASMCap improves the accuracy and power efficiency. ASMCap also embodies two hardware-friendly strategies, i.e., HDAC and TASR, to overcome the accuracy loss challenge in prior state-of-the-art approximate matching methods with negligible area cost. Results show that ASMCap outperforms the prior state-of-the-art ASM accelerators in both energy efficiency and speed with comparable accuracy.
\vspace{-0.2cm}
\section*{Acknowledgment}
\vspace{-0.1cm}
This work is supported in part by the National Key R$\&$D Program of China (No. 2019YFA0706100), in part by NSFC (U21B2030, 61934005), and in part by Tsinghua University – Daimler Greater China Ltd. Joint Institute for Sustainable Mobility. H. Zhong and Z. Chen make equal contributions.

\end{document}